\documentclass[reprint,amsmath,amssymb,superscriptaddress, prl, showpacs]{revtex4-1}
\usepackage{graphicx}
\usepackage{dcolumn}
\usepackage{bm}
\usepackage{amsmath}
\usepackage{amssymb,amsthm}
\usepackage{color}
\usepackage{epstopdf} 
\usepackage[sans]{dsfont}

\usepackage{epstopdf}
\usepackage{mathbbol,bbm}

\let\originalleft\left
\let\originalright\right
\renewcommand{\left}{\mathopen{}\mathclose\bgroup\originalleft}
\renewcommand{\right}{\aftergroup\egroup\originalright}
\newcommand{\bra}[1]{\ensuremath{\left< #1\right|}}
\newcommand{\ket}[1]{\ensuremath{\left|#1\right>}}

\newcommand{\braket}[2]{\left< #1,#2\right>}
\providecommand{\abs}[1]{\left|#1\right|}
\providecommand{\norm}[1]{||#1||}

\begin{document}
\title{Optimally designed quantum transport across disordered networks}

\author{Mattia Walschaers} 
\email{mattia@itf.fys.kuleuven.be}
\affiliation{Physikalisches Institut, Albert-Ludwigs-Universitat Freiburg, Hermann-Herder-Str. 3, D-79104 Freiburg, Germany}
\affiliation{Instituut voor Theoretische Fysica, KU Leuven, Celestijnenlaan 200D, B-3001 Heverlee, Belgium}

\author{Jorge Fernandez-de-Cossio Diaz} 
\email{jfernandez@estudiantes.fisica.uh.cu}
\affiliation{Complex System Group, Department of Theoretical Physics, University of Havana, Cuba}

\author{Roberto Mulet} 
\email{roberto.mulet@gmail.com}
\affiliation{Physikalisches Institut, Albert-Ludwigs-Universitat Freiburg, Hermann-Herder-Str. 3, D-79104 Freiburg, Germany}
\affiliation{Complex System Group, Department of Theoretical Physics, University of Havana, Cuba}

\author{Andreas Buchleitner}
\email{abu@uni-freiburg.de}
\affiliation{Physikalisches Institut, Albert-Ludwigs-Universitat Freiburg, Hermann-Herder-Str. 3, D-79104 Freiburg, Germany}

\date{\today}
\pacs{05.60.Gg, 03.65.Xp, 72.10.-d, 87.15.hj}


\begin{abstract}

We 
establish a general mechanism for highly efficient quantum transport through finite, disordered 3D networks. It 
relies on the interplay of disorder with centro-symmetry and a dominant doublet spectral structure, and can be controlled by proper tuning of 
{\em only} coarse-grained quantities. 
Photosynthetic light harvesting complexes are discussed as potential biological incarnations of this design principle.

\end{abstract}

\maketitle

In a variety of fields, ranging from quantum information \cite{Bose} to solar cell physics \cite{Nelson}, the efficient transport of quanta is of paramount importance. In realistic setups, however, one typically encounters systems which are complex in nature and only allow for a limited degree of control. Therefore it is relevant to understand which general conditions are required such that fundamental principles of quantum mechanics can be exploited to enhance transport in complex systems. At present, this question is still widely open.

Common wisdom suggests that quantum interference can enhance transport across perfectly periodic potentials \cite{Ashcroft,aharonov93}, while it tends to suppress transport in disordered systems \cite{Anderson,Modugno}. In general, multi-path quantum interference leads to erratic, large scale fluctuations of transmission probabilities when boundary conditions or other system parameters are slightly changed \cite{RichterWeidenmueller,Kramer,Madronero,Dittrich,loeck10}. These fluctuations are often indicative of the strong, non-linear coupling of few degrees of freedom, as it abounds in heavy nuclei \cite{RichterWeidenmueller}, ultra-cold many-particle dynamics \cite{rodriguez12}, strongly perturbed Rydberg systems \cite{choi04,madronero05,gurian12}, billiard geometries for photons \cite{Stoeckmann} and electrons \cite{sachrajda98}, strongly driven quantum systems \cite{carvalho}, and in large molecules \cite{Keshavamurthy,baier09,TjaartKrueger}. Often devices which transport quanta tend to avoid these fluctuations, however one may wonder whether they can be exploited.

In the present contribution our purpose is to identify design principles 
for the properties of disordered Hamiltonians that, supported by large scale fluctuations, may generate {\it quantum-enhanced} transport. These design principles are  statistically robust, in the sense that they are ``implementable'' by controlling only few coarse-grained parameters. As it is our goal to illustrate the great potential of constructive interference 
in disordered quantum systems, we choose to evade technical details of any specific implementation. We show that a collection of 
random Hamiltonians amended by only two additional constraints features high probabilities for near-to-perfect single-excitation transport across the abstract networks they can be associated with. The probability distribution of transfer efficiencies is fully controlled by the networks' electronic density of states, some {\em average} coupling matrix element, and the complex size (in terms of number of constituents), which are easily controllable e.g. in macromolecular design \cite{Kozaki1, Wasielewski, Kozaki2, Gerlich}. As a potential application, we discuss the possible role of our findings for efficient light harvesting in photosynthetic complexes. While it is not our intention to perform a detailed analysis of these complicated biological structures, we rather wish to scrutinize the relevance of the 
introduced design principles in the light of available structure data.

As a working model we consider the coherent transport of one excitation across a disordered 3D network of $N$ sites. Hilbert space is spanned by the basis states $\ket{i}$ which represent those states where the excitation is fully localized at the network's site $i$. In order to formulate a quantitative, statistical theory, we generate different realizations of disorder by sampling over $N\times N$ random Hamiltonians $H$ extracted from the Gaussian Ortogonal Ensemble (GOE) \cite{Mehta}. The matrix entries $H_{i,j}$  encode the couplings between sites $i$ and $j$. For each realization, input $\ket{\rm in}$ and output $\ket{\rm out}$ are defined as those sites with the weakest coupling $V=\min_{i\ne j}|H_{i,j}|$. Our figure of merit
  is the transfer efficiency 
\begin{equation}
\label{eq:eff}{\cal P}_H = \max_{t\in \left[0,T_R\right)} \abs{\braket{{\rm out}}{\phi(t)}}^2\, ,\, \ket{\phi(0)}=\ket{\rm in}\, ,
\end{equation} 
which quantifies a given random network's performance in terms of excitation transport from $\ket{\rm in}$ to $\ket{\rm out}$. ${\cal P}_H$ is gauged against the direct coupling $V$ between $\ket{\rm in}$ and $\ket{\rm out}$, in the absence of all intermediate sites, through the definition of the associated benchmark time scale $T_R=\pi/2V$ \cite{Scholak10,Scholak}.

Earlier studies of coupled dipoles suggested that a {\it centro-symmetric} structure of the Hamiltonian with respect to $\ket{\rm in}$ and $\ket{\rm out}$ is a valuable ingredient for perfect-state transfer in dipole-dipole networks \cite{ Christandl2,Kay,tozech}. This symmetry is defined by $JH=HJ$ and $\ket{{\rm in}}=J\ket{{\rm out}}$ \cite{footnote1}, where $J$ is the exchange matrix, $J_{i,j}=\delta_{i,N-j+1}$\cite{Cantoni}. However, for GOE Hamiltonians on which centro-symmetry is imposed, the  transfer efficiencies are still rather broadly distributed, implying that centro-symmetry alone is not sufficient for efficient state transfer. Therefore, we need to identify an additional structural element which guarantees robustness, in the sense that the transfer efficiency must not depend strongly on the specific conformation of the intermediate sites. Such a feature is also of obvious relevance for our model's applicability 
to real light harvesting complexes, which continuously undergo conformational changes (whether noisy or deterministic) on the macromolecular scale.

Intuitively, structural stability of efficient excitation transfer from $\ket{\rm in}$ to $\ket{\rm out}$ is guaranteed if both states are coupled through a {\em dominant tunneling doublet} in the spectrum. The sole role of the intermediate states is then to {\em collectively amend} the effective tunneling coupling by an energy shift $\Delta s$. If $\Delta s$, which strongly fluctuates under variations of the network conformation 
(induced by the coupling to some 
background degrees of freedom, e.g. vibrational modes of macromolecular structures \cite{Mancal}), has the proper sign, this can lead to a {\em dramatic} enhancement of the transfer efficiency. Such collective shifts induced by the coupling to random or ``chaotic" states have been investigated in the context of {\em chaos assisted tunneling (CAT)} \cite{Tomsovic,Ullmo,Zakrzewski,hensinger01}, 
and will enter as the key ingredient of the subsequent analytical description of our problem.

Given the centro-symmetry of  $H$, it can be cast, through an orthogonal transformation to the eigenbasis of the exchange operator $J$, into the block diagonal representation \cite{Cantoni}
\begin{equation}\label{eq:Hblock}H = \begin{pmatrix}
 H^+ & 0 \\
 0 & H^- \end{pmatrix}\, .
\end{equation} 
In this new form, both $H^+$ and $H^-$ are again $N/2\times N/2$ GOE matrices, {\it i.e.} the elements $H^{\pm}_{i,j}$ are sampled from a Gaussian distribution with zero mean and variance $(1+\delta_{i,j}) 2 \xi^2/N$. 

Since two of the eigenvectors of $J$ have the form $\ket{\pm}=\frac{1}{\sqrt{2}}\left(\ket{{\rm in}} \pm \ket{{\rm out}}\right)\, $
we now additionally assume (see above) that $\ket{+}$ and $\ket{-}$ form a dominant doublet, such that they are both close to eigenstates $\ket{\tilde{+}}$ and $\ket{\tilde{-}}$ of  $H^+$ and $H^-$, respectively \cite{footnote3}. It is then useful to write the Hamiltonian (\ref{eq:Hblock}) as
\begin{equation}
\label{eq:Matrix}
H= \begin{pmatrix} E+V & \bra{\mathcal{V}^+}& &  \\ \ket{\mathcal{V}^+} & H^+_{sub}& & \\
& & E-V & \bra{\mathcal{V}^-}\\ & & \ket{\mathcal{V}^-} &H_{sub}^- \end{pmatrix}\, ,
\end{equation} 
which makes the definition of rows and columns which relate to $\ket{+}$ and $\ket{-}$ explicit. 
From the definition of $\ket{\pm}$ it is easy to see that $\bra{\pm}H\ket{\pm}=E\pm V$. $\ket{\mathcal{V}^{\pm}}$ contains the (Gaussian distributed) 
couplings of the dominant doublet states $\ket{\pm}$ to the remainder of the system. 

Due to the dominant doublet assumption,$\abs{\braket{\tilde{\pm}}{\pm}}^2 > \alpha \approx 1\, ,$
the norm $\norm{\mathcal{V}^{\pm}}$ of the coupling is  small and, under this condition, perturbation theory guarantees that $E \pm V$ in (\ref{eq:Matrix}) are eigenvalues of $H$, up to some perturbative correction term $s^{\pm}$. The explicit expression for the transfer efficiency is then dominated by those terms associated with $\ket{\tilde{\pm}}$, leading to the estimate
\begin{equation}
\label{eq:finalEff}
\mathcal{P}_H > \max_{t \in [0.T_R)}  \frac{\alpha^2}{4}\abs{ e^{-i t (E + V + s^+)}  - e^{-i t  (E-V+s^-)}}^2
\end{equation} 
where $\quad  s^{\pm}= \sum_i \frac{\abs{\braket{\mathcal{V}^{\pm}}{\psi^{\pm}_i}}^2}{E\pm V - e^{\pm}_i}$ and $\ket{\psi^{\pm}_i}$ and $e^{\pm}_i$ are the eigenvectors and eigenvalues of $H^{\pm}_{sub}$. From (\ref{eq:finalEff}) it is clear that the efficiency is large, $\mathcal{P}_H > \alpha^2$, if  $t=\pi/\abs{2V+\Delta s}$, $\Delta s=s^+-s^-$, is smaller than $T_R$, and we can interpret $\abs{2V+\Delta s}$ as an effective tunneling rate. All realizations for which we obtain $T_R/t > 1$ have efficient transport which is faster than the direct coupling between $\ket{\rm in}$ and $\ket{\rm out}$.  Note that the dominant doublet assumption alone does not guarantee this latter condition, rather this is a fundamental consequence of the strong fluctuations that arise due to the disorder. This may be induced, for example, by the coupling of some complex background degrees of freedom, such as vibrational modes \cite{Mancal}. Only a sufficiently broad distribution of $\Delta s$ guarantees 
that efficient transfer can {\em always} be achieved, even if the direct coupling $V$ between input and output site {\em vanishes} \cite{Tomsovic}.  Thus, despite weakly coupled, the presence of the intermediate, random sites of the network as represented by $H_{\rm sub}^\pm$ is absolutely crucial to achieve efficient transport.

For fixed $E$ and $V$ the distribution of $\Delta s$ was already derived within the context of CAT \cite{Ullmo,Zakrzewski} and in \cite{Fyodorov}, and it turned out that for large $N$ it is a Cauchy distribution. In our present problem, $E$ and $V$ are themselves stochastic variables, and, therefore, should be averaged over. Since the integrations over $E$ and $V$ are dominated by their mean values, given by $\overline{V}\approx 2\pi \sqrt{2}\xi e^{-1} N^{-3/2}$ and $\overline{E}=0$, a lengthy but straightforward calculation shows that the probability distribution of $T_R(2V+\Delta s)/\pi =T_R/t$ is given by
\begin{equation}
\label{eq:Dist}
P\left(\frac{T_R}{t}=x\right)=\frac{1}{\pi}\left(\frac{s_0}{{s_0}^2+(1+x_0+x)^2 }+\frac{s_0}{{s_0}^2+(1+x_0-x)^2 }\right)\, ,
\end{equation}
with $s_0=\frac{\overline{\norm{\mathcal{V}}^2} N e \left(1-2/N\right)^{1/2}}{4 \pi \xi^2}\,$, $\quad x_0=\frac{\overline{\norm{\mathcal{V}}^2}}{2\xi^2}$, and $\overline{\norm{\mathcal{V}}^2}$ the expectation value of $\norm{\mathcal{V}^{\pm}}^2$ for 
all realizations where the dominant doublet assumption holds. 

The distribution (\ref{eq:Dist}) depends on only two {\em coarse grained} parameters: $\xi$ characterizes the spectral density of the eigenstates of $H_{sub}^+$ and $H_{sub}^-$, while $\overline{\norm{\mathcal{V}}^2}$ measures the average coupling strength of the dominant doublet to these states. It therefore cannot be emphasized enough that, within the picture here elaborated, the transport properties of the problem {\em do not} depend on the specificities of the Hamiltonian, or the intermediate electronic states of the network.


To validate our theoretical model by numerical simulations, we generate many GOE Hamiltonians with the additional constraint of centro-symmetry with respect to $\ket{\rm in}$ and $\ket{\rm out}$. For each of these Hamiltonians the existence of  a dominant doublet is assessed by inspection of its eigenvectors and by verifying that, for some $\ket{\tilde{\pm}}$, the condition $\abs{\braket{\tilde{\pm}}{\pm}}^2 > \alpha$ holds. This post-selection defines the statistical ensemble which we expect to satisfy (\ref{eq:Dist}). The quantity $\mathcal{P}_H$ is then obtained by numerical propagation of the quantum dynamics generated by the Hamiltonian, and $t$ is defined as the earliest point in time for which $\abs{\braket{{\rm out}}{\phi(t)}}^2=\mathcal{P}_H$.

Fig.~1 compares numerical results and analytical prediction (\ref{eq:Dist}). It must be noticed that there are no free parameters involved in this plot: The values $\xi=2$, $\alpha>0.95$ and $\overline{\norm{\mathcal{V}}^2}=0.311962$ which enter (5) are either given a priori or directly extracted from the statistical characterization of the numerically generated sample Hamiltonians. 
\begin{figure}[!htb]
\centering
\includegraphics[scale=0.6]{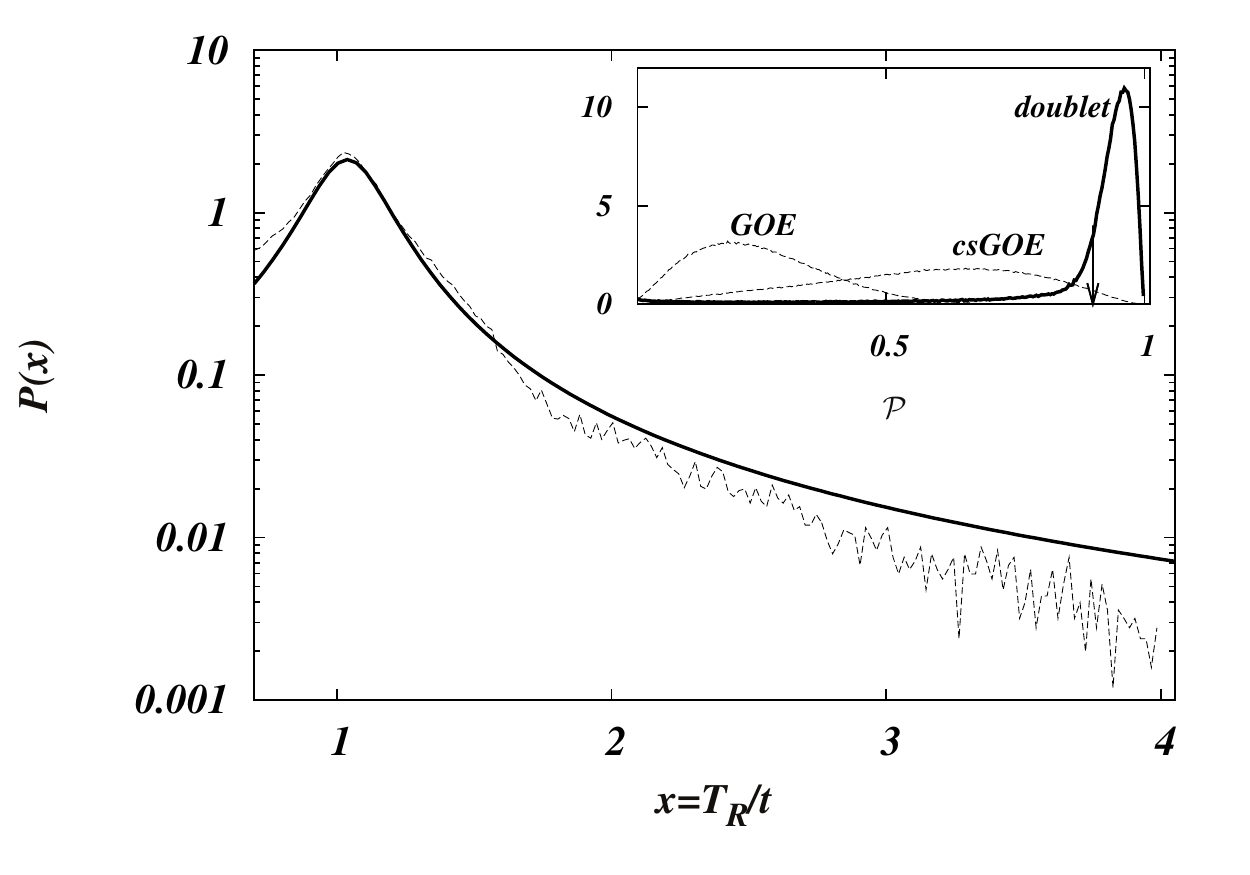}

\caption{Histogram of the numerically simulated inverse transfer time $T_R/t$ (dashed line), compared to the theoretical prediction (\ref{eq:Dist}) (full line), for density of states $\xi=2$, $N=10$ network sites, and a dominant doublet condition $\alpha=0.95$. $\overline{\norm{\mathcal{V}}^2}=0.311962$ is directly extracted from the numerical sampling. Simulations were performed in the time window $[0,1.7 T_R]$. Inset: Histograms of the transfer efficiencies $\mathcal{P}_H$ for three different matrix ensembles, fixed  density of states $\xi=2$, $N=10$ network sites. As the ensemble is constrained from GOE to centro-symmetric and to centro-symmetric with dominant doublet condition $\alpha=0.95$, the average efficiencies are dramatically enhanced.} 
\label{fig:time}
\end{figure}

Clearly, the majority of realizations have $t$ smaller than the time scale set by the direct coupling $V$ between $\ket{\rm in}$ and $\ket{\rm out}$. The fat algebraic tail of the Cauchy distribution for $t\ll T_R$ guarantees that realizations with very fast transport are abundant in the sense that they are not exponentially unlikely. 

For fixed $\alpha = 0.95$, our model predicts efficiencies larger than $\mathcal{P}_H > \alpha^2\approx 0.9$. This is indeed observed in the simulations. The inset in Fig.~1 shows the probability distribution of the efficiencies, which is sharply peaked above $\mathcal{P}_H > 0.9$ (see arrow). Comparison, in the same figure, with the probability distributions of the efficiencies of  centro-symmetric GOE matrices without the doublet constraint, and of general GOE matrices, respectively, shows that in both cases the average efficiency is significantly lower than for those centro-symmetric Hamiltonians which exhibit the
additional design element of a dominant doublet. 

A remarkable asset of this transport mechanism is its robustness under different realizations of disorder, which,  in the context of networks, refers to different configurations of the intermediate sites (represented by the random matrices $H_{\rm sub}^{\pm}$ in (\ref{eq:Matrix})). In the light of the recent debate on the potential role and unexpected robustness of quantum coherence in photosynthetic harvesting of the sunlight's energy \cite{Mancal}, one may wonder whether the proposed design principles are implemented by Nature.
Indeed, some of the light harvesting complexes which are 
hardwired in  bacteria or plants, such as the FMO complex of green sulfur bacteria \cite{Trorund,Schmidt11,greg+abu12}  
exhibit an apparently disordered, network-like structure, and appear to be optimized for efficient transport. 

It is therefore suggestive to test the hypothesis that centro-symmetry and dominant doublet are compatible with the available structure 
data \cite{Trorund,Schmidt11}. 
For this purpose, 
we fix the spatial position of the FMO's constituent BChl{\it a} molecules as given in the literature \cite{Trorund} (see Table I, Supplementary Material), 
and only allow the orientation of the dipoles associated with each of the BChl{\it a}'s to vary. Furthermore, we neglect on-site energy
shifts induced by the coupling to background degrees of freedom, i.e., all on-site energies are assumed to be identical. 
Apart from a possible limitation of the maximally achievable transfer efficiencies to values smaller than $100\ \%$ (similar to the limitation of the 
maximum  transfer amplitude by the bias in an asymmetric double-well potential) this does not affect the central features of our transport scenario \cite{tozech}. 

\begin{figure}[!htb]
\centering
\includegraphics[scale=0.5]{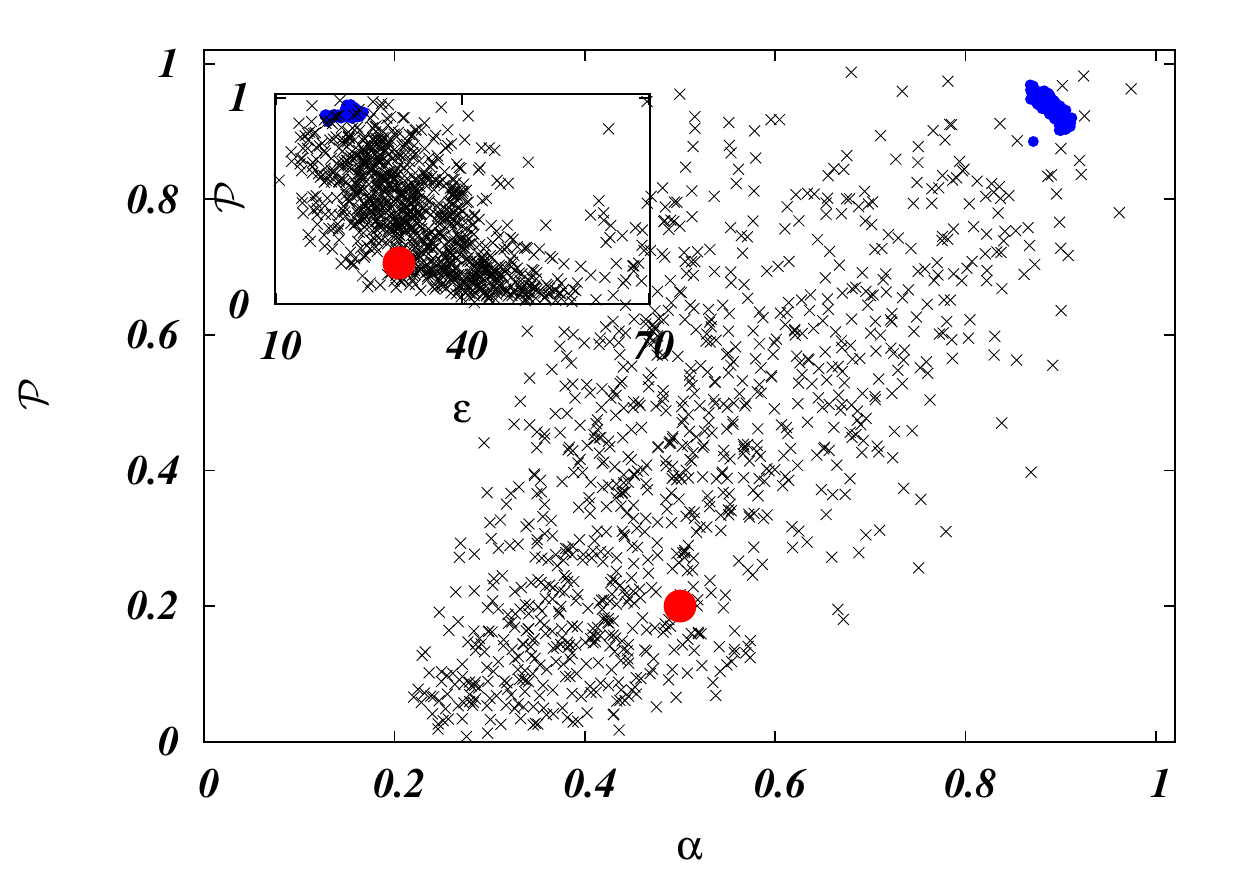}
\caption{Scatter plots of transfer efficiency $\mathcal{P}$, eq.~(1), versus centro-symmetry $\alpha$ (main figure) and dominant doublet strength $\epsilon$ (inset). Evolutionary optimisation as achieved by a genetic algorithm is indicated by the filled blue circles, upon seeding of the algorithm with the documented FMO structure (filled red circles) as listed in Tables I and II of Supplementary Material  \cite{Trorund,Schmidt11}. 
The unambiguous and synchronous attraction towards more efficient, centro-symmetric dipole orientations with large doublet strengths is to be compared to a benchmark ensemble generated by the algorithm when seeded with randomised dipole orientations (crosses).}
\label{fig:FMO}
\end{figure}

Given the spatial positions of the dipoles, the inter-site 
dipole-dipole coupling matrix elements 
$H_{i,j}$ are determined by their relative 
orientations  \cite{Schmidt11} (Table II, Supplementary Material). 
To certify the relevance of our dominant doublet picture elaborated above for abstract, disordered networks, we now ask the question how close the documented FMO conformations
are to optimal conformations in the above sense. 
To give an answer to this question, we use the tabulated FMO data to seed a genetic algorithm with the transfer efficiency (\ref{eq:eff}) as target function, and only allow for 
variations of the intermediate sites' dipole orientations (variations of the coupling to and between 
the intermediate sites generate the nontrivial and crucial statistics 
of the level shifts $\Delta s$ in the CAT scenario that underlies our analysis).
We then 
correlate the thus achieved 
optimal transfer efficiencies with the optimal networks' centro-symmetry quantifiers \cite{tozech},  $\epsilon =\frac{1}{N}\min_S ||H - J^{-1}HJ||$
(where the minimisation runs over all permutations of the intermediate network sites $2,\dots ,N-1$, and the Hilbert-Schmidt norm \cite{Reed} is employed), and the dominant doublet strengths $\alpha$ (here
defined as the minimum of $\abs{\braket{\tilde{+}}{+}}^2$ and $\abs{\braket{\tilde{-}}{-}}^2$).  
These 
results are benchmarked against optimisation results seeded by random orientations of the dipoles, and illustrated in 
Fig.~\ref{fig:FMO}. Filled blue circles represent the results delivered by the genetic algorithm when launched in the vicinity of the documented FMO structure -- which itself exhibits (poor) efficiency, doublet strength and centro-symmetry as represented by the red filled circles in both plots. The synchronous trend towards significantly 
enhanced efficiencies, centro-symmetries and doublet strengths is {\em unambiguous and in stark contrast} 
to the benchmark ensemble represented by crosses
in the plot, which also reflect some correlation between efficiency, centro-symmetry and doublet strength, but lack the essentially deterministic
attraction towards optimal performance which manifests in the FMO's vicinity. 

\begin{figure}
\centering
\includegraphics[scale=0.25]{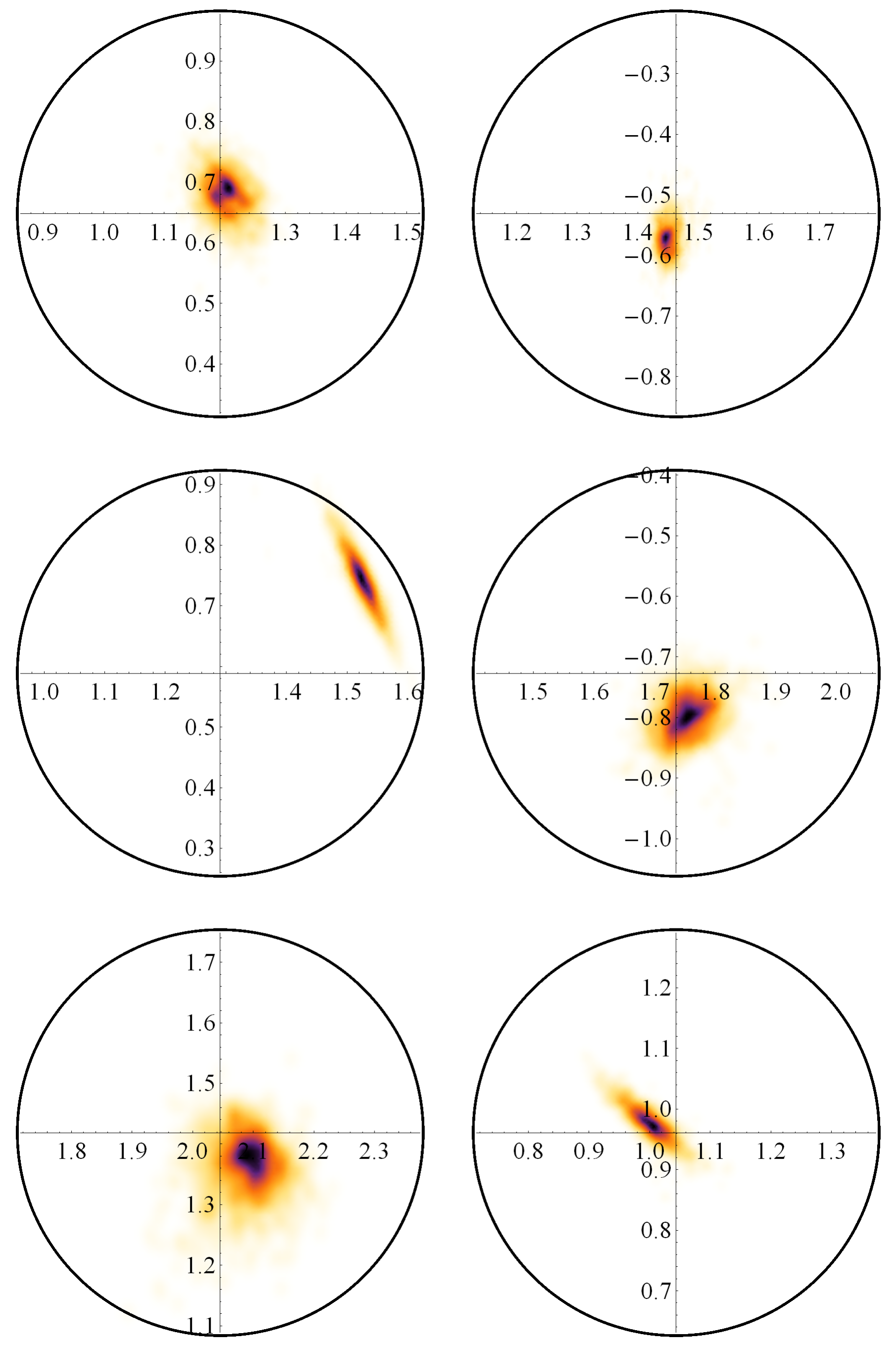}
\caption{(Linearly) Grey scaled probability density of the genetically optimised FMO dipole orientations, in spherical coordinates ($\phi,\theta$) (in radian). Dipoles 1,2,4,5,6,7 are listed from left to right and top to bottom, and the experimental dipole orientation extracted from Table II (Supplementary Material) defines the origin of each plot. Dipoles 8 (input) and 3 (output) are not shown since we keep their orientations fixed during optimization.}
\label{fig:Angles}
\end{figure}

On top of this evolutionary attraction towards optimal performance in the FMO neighbourhood, in response to our above question, 
the dipole orientations which result from evolutionary optimisation are indeed very close to the dipole orientations as given by the experimental data: Fig. \ref{fig:Angles} depicts the probability densities for the relative positions of the {\em optimal} dipole orientations at each of the intermediate BChl{\it a} sites, in ($\phi,\theta$) spherical coordinates with respect to the tabulated orientations 
which define the origin of each plot.  In the worst case, the average orientation of dipole 4 deviates by less than $20\%$ from the experimental data. All the other optimised dipole orientations deviate by less than $7\%$. Therefore the documented FMO dipole 
structure has a design close to optimal with respect to the abstract design principles which we introduced above. 

In summary, we described a general mechanism that gives rise to fast and efficient quantum transport in finite, 3D 
disordered systems. The mechanism rests on two crucial design principles: The {\em centro-symmetry} of the underlying Hamiltonian, which guarantees a natural block diagonal representation, and the existence of a {\em dominant doublet}, that ensures that the coupling to random/chaotic states (provided, e.g., by the intermediate sites of a molecular network) can efficiently assist the transport in a robust way. The statistics of the transfer efficiencies and times, as shown in Fig.~1,  then only depend on the intermediate network sites' density of states $\xi$, and on the average coupling strength $\overline{\norm{\mathcal{V}}^2}$ of the in- and output-sites to the network. While the former can be controlled e.g. by the packing of the intermediate network sites, the latter should be easily controllable by fixing -- e.g. through a protein scaffold \cite{scholes11} -- the average distance from input and output to the intermediate sites. Within this perspective, robust and efficient transport across complex quantum networks may be  achieved by optimally designing {\em not} one single network conformation, but rather a {\em suitable statistical distribution}, fixed by the density of states and some average coupling strength alone. 

\textbf{Acknowledgments:} R.M. acknowledges support by the Alexander von  Humboldt Stiftung. M.W. and A.B. are grateful 
for funding within the DFG Research Unit  760, and for support through the EU COST Action MP1006. M.W. acknowledges partial funding by Belgian Interuniversity Attraction Poles Programme P6/02 and FWO Vlaanderen project G040710N.


\onecolumngrid

\section{Supplementary Material}

\subsection{Genetic algorithm}

The optimization algorithm resembles the one used in \cite{tozech}, and is seeded with the spatial coordinates of FMO (Table I), together with 
the associated 8 dipole moments $\vec{d}_i$, $i = 1,\ldots ,8$ (Table II), that determine the off-diagonal elements of the Hamiltonian $H$, through a dipole-dipole approximation \cite{Schmidt11}. When numbering the dipoles we follow the standard notation \cite{Trorund}.

\begin{enumerate}
\item Each one of the intermediate ($i=1,2,4,5,6,7$) sites' dipole moments' orientations is subject to 100 
random perturbations, to generate new dipoles $\vec{d}_{i}^{{\rm new}}$ from the old ones $\vec{d}_{i}^{{\rm old}}$,
according to the following procedure:
\begin{eqnarray*}
\vec{b}_{i} & = & \vec{d}_{i}^{{\rm old}}+r_{i}\vec{n}_{i}\\
\vec{d}_{i}^{{\rm new}} & = & \vec{b}_{i}/\left|\vec{b}_{i}\right| \, .
\end{eqnarray*}
Here $r_{i}$ is a random Gaussian variable with zero mean and standard
deviation $\sigma$ (initially set to $\sigma =0.005$), and $\vec{n}_{i}$
is a randomly oriented unit vector generated with the GSL (GNU Scientific Library)
routine \emph{gsl\_ran\_dir\_3d}, with the additional condition $\left|\vec{b}_{i}\right|\geq 0.1$.
\item These new dipole configurations define 100 new Hamiltonians $H$ and, correspondingly, 100 new, different values of 
the quantum transfer efficiency ${\cal P}$, from input site 8 to output site 3. 
\item That configuration which mediates the largest efficiency defines the new set of dipole moments $\vec{d}_i$.
\item We repeat steps 1 to 3 above, with the new $\vec{d}_i$, and reduce $\sigma$ to $\sigma/k$, 
in the $k$th iteration.
\item The algorithm stops when ${\cal P}=1-x$, with $x<0.01$, or when $k=100$.
 \end{enumerate}
When seeded with the experimental FMO data, the algorithm generates efficient configurations very rapidly. In this case we typically reach convergence in less than 20 iterations. However, when the algorithm is seeded with a random dipole configuration, there seems 
no tendency to converge to high efficiencies. In this case, the algorithm saturates at low values of ${\cal P}$ (see Figure 2), after less than 50 iterations.

\subsection{Coordinates and dipole moments of the FMO structure}

\begin{table*}[!htb]\centering
{\small
\begin{tabular}{cccc}
Site & $x$ & $y$ & $z$ \\
\hline
1 & 26.51 & 2.597 & -11.349 \\
2 & 15.607 & -1.517 & -17.246\\
3 & 3.389 & -13.614 & -13.851\\
4 & 6.678 & -20.848 & -6.036\\
5 & 19.378 & -18.571 & -1.076\\
6 & 21.834 & -7.175 & 0.634\\
7 &  10.274 & -8.207 & -5.544\\
8 &  21.766 & 13.748 & -7.718\\
\hline
\end{tabular}
}
\caption{Spatial coordinates of the BChl{\it a} molecules of the FMO (in Angstroms), extracted from file 3ENI.pdb1 in the Protein Data Bank \cite{Trorund}.} 
\end{table*}

Given the spatial positions of the (unit-length) dipoles, 
the inter-site dipole-dipole coupling matrix elements $H_{i,j}$ are determined \cite{Schmidt11} by their relative 
orientations, listed in Table II \cite{Trorund}. 

\begin{table*}[!htb]\centering
{\small
\begin{tabular}{cccc}
Site & $S_x$ & $S_y$ & $S_z$ \\
\hline
1 & 0.741006 & 0.560602 & 0.369644\\
2 &  0.857141 & -0.503776 & 0.107329\\
3 &  0.197121 & -0.95741 & 0.210971\\
4 &  0.760508 & 0.593481&  0.263453\\
5 &  0.736925 & -0.655762& -0.164065\\
6 &  0.135017 & 0.879218& -0.456887\\
7 &  0.495115 & 0.708341& 0.503105\\
8 &  0.553292 & 0.138385& -0.821412\\
\hline
\end{tabular}
}
\caption{Normalized dipole components of the BChl{\it a} molecules of the FMO complex, extracted from
file 3ENI.pdb1 in the Protein Data Bank  \cite{Trorund}.}
\end{table*}


\begin{thebibliography}{99}
\bibitem{footnote3} Note that the dominant doublet assumption is much more natural 
\cite{Tomsovic} given centro-symmetry than in its absence: While, in the former case, the dominant doublet is an element 
of the natural basis given by the eigenstates of $J$, it is an ad hoc assumption without such discrete symmetry of the underlying Hamiltonian.
Accordingly, it can be shown that the probability density of dominant doublets without centro-symmetry is quadratically suppressed as compared to the centro-symmetric case.
\end{thebibliography}
\end{document}